\documentclass[preprint,review,12pt]{elsarticle}
\usepackage{amssymb}
\journal{Chaos, Solitons and Fractals}
\begin{document}

\begin{frontmatter}

\title{Analytical solution of SEIR model describing the free spread of the COVID-19 pandemic}

\author{Nicola Piovella}

\address{Dipartimento di Fisica "Aldo Pontremoli", Universit\`a degli Studi di Milano, Via Celoria 16, Milano I-20133, Italy}

\begin{abstract}
We analytically study the SEIR (Susceptible Exposed Infectious Removed)  epidemic model. The aim is to provide simple analytical expressions for the peak and asymptotic values and their characteristic times of the populations affected by the COVID-19 pandemic.
\end{abstract}


\end{frontmatter}

\section{Introduction}

The COVID-19 outbreak has motivated  a large number of numerical studies using epidemiology models \cite{rodriguez2020clinical,yang2020modified}. A commonly used model is the Susceptible–-Exposed–-Infected–-Removed (SEIR) model  \cite{kermack1927contribution}. This model is formulated as a system of nonlinear ordinary differential equations, for which no exact analytic solution has yet been found. For this reason, most of the recent works focus on the numerical analysis of statistical ensembles of initial data for these equations. However, due to the uncertainty and often unreliability of the clinical data, the prediction about the real evolution of the epidemic is rather difficult, if not impossible \cite{faranda2020modelling,bertozzi2020challenges}. On the other hand, the SEIR epidemic model provides a deterministic evolution for some given initial state. Therefore, the aim of this work is to provide simple expressions of the main characteristics of the population of individuals that have been in contact with the disease, as of instance the peak of the infected population and the time after which it occurs, the final number of individuals who have contracted the disease and the temporal shape of the infectious population's curves. These analytical expressions can become useful through their application to the COVID-19, to obtain fundamental parameters as the reproduction number $r$ and the epidemic starting time.

The paper is organized as follow, In sec. II we recall the SEIR model; in sec. III we study the linear regime with the exponential growing and decaying evolution, depending on the reproduction number $r$; in sec. IV we investigate the nonlinear regime in the free spread evolution with $r>1$. We approximate the exact model of equations by a reduced model where the decaying mode is adiabatically eliminated. This reduced model allows to obtain analytical results which have been seen to be in good agreement with the exact numerical solution. Sec. V summarizes the results and draws the conclusions.

\section{The SEIR model}

We used the susceptible–-exposed–-infected–-removed (SEIR) compartment model \cite{kermack1927contribution,anderson1992infectious,pastor2015epidemic,zhou2020preliminary} to characterize the early spreading of COVID-19, where each individual could be in one of the following states: susceptible ($S$), exposed ($E$, being infected but without infectiousness),
infected ($I$, with infectiousness), recovered ($R$) and dead ($D$). At later times a susceptible individual in the state $S$ would turn to be an individual in the exposed state $E$ with a rate $r/\tau_I$, where $r$  is the reproduction number (i.e. the average number of infected people generated by each infected person during the desease) and $\tau_I=1/\gamma_2$ is the average time in the infected state $I$. An exposed individual in the state $E$ becomes infected, i.e. in the state $I$ in an average time $\tau_E=1/\gamma_1$. Then the infected individual is removed from the total population with the rate $\gamma_2$ either by recovering ($R$) or dying ($D$) with 
a mean case fatality proportion $p$. The dynamical process of SEIR is described by the following set of equations:
\begin{eqnarray}
\dot S &=& -r\gamma_2\left(\frac{S}{N}\right)I,\label{eq:S} \\
\dot E &=& r\gamma_2\left(\frac{S}{N}\right)I-\gamma_1E,\label{eq:E}\\
\dot I &=& \gamma_1 E-\gamma_2 I,\label{eq:I}\\
\dot R &=& (1-p)\gamma_2 I,\label{eq:R}\\
\dot D &=& p\gamma_2 I.\label{eq:D}
\end{eqnarray}
Here $S(t)$, $E(t)$, $I(t)$, $R(t)$ and $D(t)$ respectively represent the number of individuals in the susceptible, exposed, infectious, recovered and death states at time $t$ and $N$ is the total number of individuals in the system such that $N(t)=S(t)+E(t)+I(t)+R(t)$.
Finally, the cumulative population $C$ is
\begin{equation}
C=E+I+R+D,
\end{equation}
equal to the total population of individuals who have contracted the infection.

\section{Linear regime}
If $E(t),I(t),R(t)\ll N(t)$, then the susceptible population $S$ can be approximated by the total population $N$ (i.e. $S\sim N$) and the equations for the exposed and infected population are linear:
\begin{eqnarray}
\dot E &=& r\gamma_2I-\gamma_1E\label{eq:E:lin}\\
\dot I &=& \gamma_1E-\gamma_2I\label{eq:I:lin}
\end{eqnarray}
 
\subsection{General solution of the linear equations}
Introducing the Laplace transforms
\begin{eqnarray*}
\tilde E(\lambda) &=& \int_0^\infty E(t)e^{-\lambda t}dt\\
\tilde I(\lambda) &=& \int_0^\infty I(t)e^{-\lambda t}dt
\end{eqnarray*}
with $\mathrm{Re} \lambda>0$, Eqs.(\ref{eq:E:lin}) and (\ref{eq:I:lin}) becomes
\begin{equation}
\left(%
\begin{array}{cc}
  \lambda+\gamma_1 & -r\gamma_2 \\
   -\gamma_1 & \lambda+ \gamma_2 \\
\end{array}%
\right)\left(
\begin{array}{c}
  \tilde E \\
  \tilde I \\
\end{array}
\right)=
\left(
\begin{array}{c}
  E(0) \\
  I(0) \\
\end{array}
\right)
\end{equation}
where $E(0)$ and $I(0)$ are the initial conditions.
The eigenvalues $\lambda$ are solution of
\begin{equation}
\left|%
\begin{array}{cc}
  \lambda+\gamma_1 & -r\gamma_2 \\
  -\gamma_1 & \lambda+\gamma_2 \\
\end{array}%
\right|=0
\end{equation}
giving
\begin{equation}
\det(\lambda)=\lambda^2+(\gamma_1+\gamma_2)\lambda+\gamma_1\gamma_2(1-r)=0
\end{equation}
with solutions
\begin{equation}
\lambda_{\pm}=-\frac{\gamma_1+\gamma_2}{2}
\pm \frac{1}{2}\sqrt{\Delta}\label{eigen}
\end{equation}
where 
\begin{equation}
\Delta=(\gamma_1+\gamma_2)^2+4\gamma_1\gamma_2(r-1)=
(\gamma_1-\gamma_2)^2+4r\gamma_1\gamma_2
\end{equation}
Since $\Delta>0$ the eigenvalues are real. Depending on $r$, we distinguish three cases:
\begin{description}
\item[(a)]  If $r>1$ then $\sqrt{\Delta}>\gamma_1+\gamma_2$, so that $\lambda_+>0$ and $\lambda_-<0$. The solution grows exponentially (explosive regime);
\item[(b)] If $r<1$ then $\sqrt{\Delta}<\gamma_1+\gamma_2$, so that both $\lambda_+<0$ and $\lambda_-<0$. The solution decays exponentially (relaxation regime);
\item[(c)] If $r=1$ then $\sqrt{\Delta}=\gamma_1+\gamma_2$, so that $\lambda_+=0$ and $\lambda_-=-(\gamma_1+\gamma_2)$. The solution remains partially constant (marginally stable regime).
\end{description}
For the cases (a) and (b) the solution is
\begin{eqnarray}
E(t) &=&\frac{1}{\sqrt{\Delta}}\left\{
\sqrt{\Delta}E(0)\cosh(\sqrt{\Delta}t/2)
+[(\gamma_2-\gamma_1)E(0)+2r\gamma_2I(0)]\sinh(\sqrt{\Delta}t/2)\right\}e^{-(\gamma_1+\gamma_2)t/2}\\
I(t) &=&\frac{1}{\sqrt{\Delta}}\left\{
\sqrt{\Delta}I(0)\cosh(\sqrt{\Delta}t/2)
+[(\gamma_1-\gamma_2)I(0)+2\gamma_1 E(0)]\sinh(\sqrt{\Delta}t/2)
\right\}
e^{-(\gamma_1+\gamma_2)t/2}
\end{eqnarray}
whereas in the case (c) ($r=1$) the solution is
\begin{eqnarray}
E(t) &=& \frac{1}{2}[E(0)+I(0)]+\frac{1}{2}
[E(0)-I(0)]e^{-(\gamma_1+\gamma_2)t/2}\\
I(t) &=& \frac{1}{2}[E(0)+I(0)]-\frac{1}{2}
[E(0)-I(0)]e^{-(\gamma_1+\gamma_2)t/2}.
\end{eqnarray}

\subsection{Analysis}

The only parameter which can be controlled by confinement measures is the reproduction number $r$. In the following we assume that for COVID-19 the characteristic times are $\tau_E=3.69$ days and $\tau_I=3.48$ days \cite{li2020substantial}. We consider the time evolution of the population $E$ and $I$ for $r>1$, $r=1$ and $r<1$, corresponding to the explosive, marginally stable and relaxation regimes, respectively.

\subsubsection{Explosive regime}

For $r>1$ and $\lambda_+t\gg 1$, 
\begin{eqnarray}
E(t) &=&\frac{1}{2}\left\{
E(0)+\frac{1}{\sqrt{\Delta}}[(\gamma_2-\gamma_1)E(0)+2r\gamma_2I(0)]\right\}e^{\lambda_+ t}\\
I(t) &=&\frac{1}{2}\left\{
I(0)+\frac{1}{\sqrt{\Delta}}[(\gamma_1-\gamma_2)I(0)+2\gamma_1 E(0)]
\right\}
e^{\lambda_+ t}
\end{eqnarray}
with $\lambda_+>0$.

\subsubsection{Marginally stable regime}

When $r=1$, in the asymptotic limit $t\gg \tau_E,\tau_I$, $E$ and $I$ are constant,
\begin{equation}
E=I=\frac{1}{2}[E(0)+I(0)]
\end{equation}
and the death population grows linearly in time
\begin{eqnarray}
D(t) &=&D(0)+\frac{p\gamma_2}{\gamma_1+\gamma_2}[I(0)-E(0)]+\frac{1}{2}p\gamma_2[E(0)+I(0)] t
\end{eqnarray}
where $D(0)$, $E(0)$ and $I(0)$ are the values taken at time when $r$ starts to be $r=1$.

\subsubsection{Relaxation regime}

\begin{figure} 
\begin{center}
\includegraphics[width=15cm]{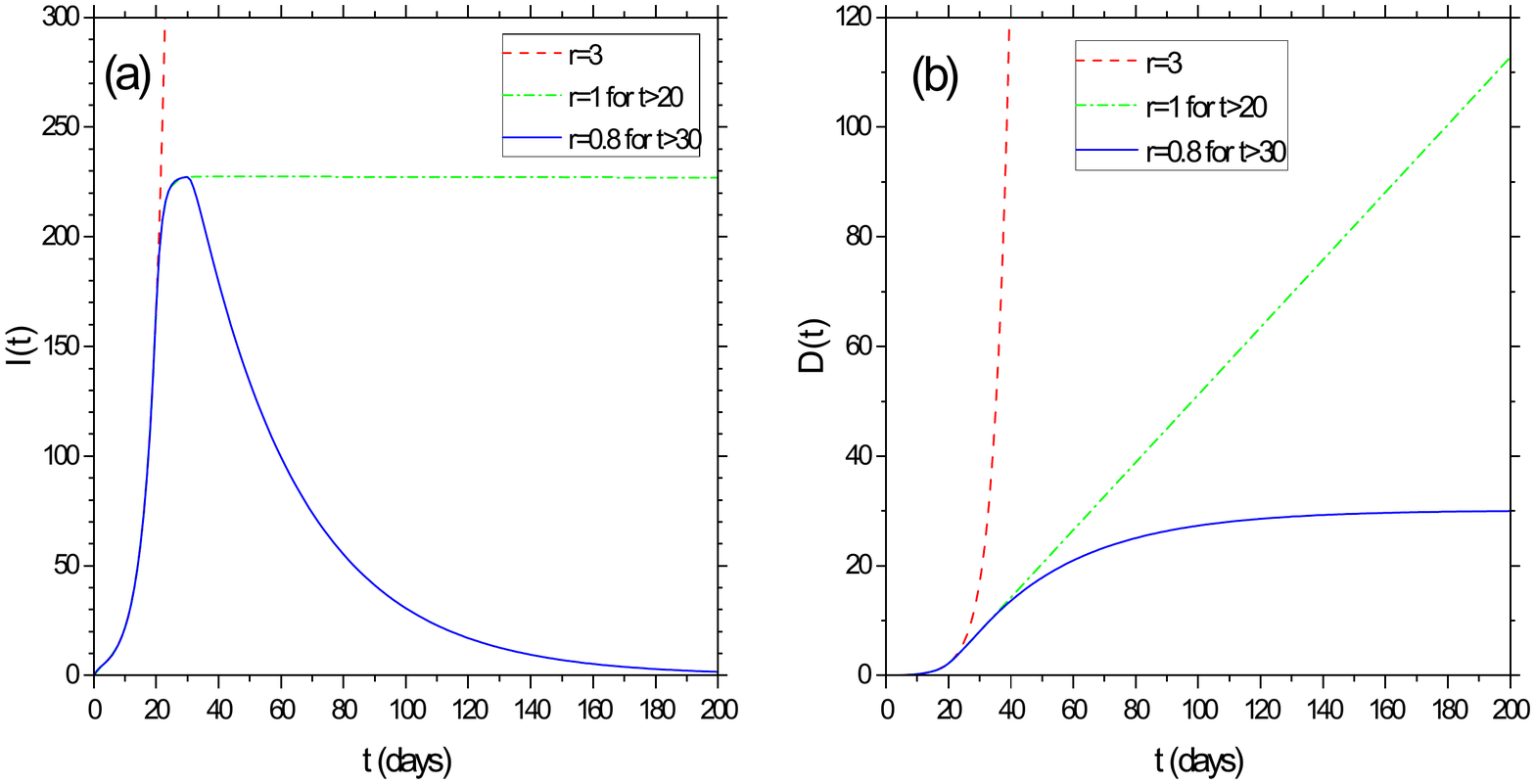}
\caption{Evolution of $I(t)$ and $D(t)$ with $r=3$ (red line), $r=1$ after $t=20$ days (green line) and $r=0.8$ after $t=30$ days (blue line). Initial conditions: $E(0)=10$. $I(0)=0$, $N(0)=6.e7$; $p=0.01$.}
\label{ID:r}
\end{center}
\end{figure}

When $r<1$, $\lambda_+$ is negative and $E$ and $I$ tend to zero, whereas $D$ tends to the following constant value,
\begin{eqnarray}
D(\infty) &=&D(0)+\frac{p}{2\gamma_2(1-r)}\left\{
[\gamma_1+\gamma_2(1+2r)]I(0)+(\gamma_2-\gamma_1)E(0)
\right\}
\end{eqnarray}
where $D(0)$, $E(0)$ and $I(0)$ are the values taken at time when $r$ starts to be $r<1$.
Fig. \ref{ID:r} shows a typical temporal evolution of $I(t)$ and $D(t)$ starting with $r>1$, then subsequently changed to $r=1$ and later on to a value $r<1$. The regime is linear (i.e. with $E,I\ll N$), the initial values are $E(0)=10$ and $I(0)=0$ and $p=0.01$. The red dashed line is for $r=3$ (explosive regime). The green dashed-dotted line is for $r$ changed from $r=3$ to $r=1$ at $t=20$ (marginally stable regime) and the blue solid line is for $r=3$ until $t=20$, then $r=1$ between $t=20$ and $t=30$ and finally $r=0.8$ for $t>30$ (relaxation regime). Notice the asymmetry of the curve of $I(t)$ due to the different growing and decaying rates.

\section{Nonlinear regime}

In the following, we investigate the nonlinear regime with a constant reproduction number $r>1$. This corresponds to a free spread of the infection, with an initial exponential growth of the exposed population $E$, and so also of $I$ and $D$. The exponential growth stops when susceptible population $S$ becomes sensibly less then the total number $N$ of the individuals. This regime is similar to the saturation in a single-mode laser, where steady-state is reached when the gain of emitted photons equals the losses by the cavity \cite{haken1970laser}.
Notice that
\begin{equation}
E+I+S+R+D=N_0
\end{equation}
is a constant of motion and $N(t)=N_0-D(t)$. However, if $p\ll 1$ we always have $D\ll N_0$, so that with a good approximation we can approximate $N$ by $N_0$. Introducing the removed population $Q=R+D$, we can eliminate $S=N_0-(E+I+Q)$ using the constant of motion and obtain
\begin{eqnarray}
\dot E &=& r\gamma_2\left(1-\frac{E+I+Q}{N_0}\right)I-\gamma_1 E\label{eq:ES:2bis}\\
\dot I &=& \gamma_1 E-\gamma_2 I\label{eq:I:3}\\
\dot Q &=& \gamma_2 I
\end{eqnarray}
We normalize the variables by $N_0$ defining $x=E/N_0$, $y=I/N_0$ and $z=C/N_0$ where $C=E+I+Q$ is the cumulative population, i.e. the total number of individuals who have contracted the infection. Then the equations become
\begin{eqnarray}
\dot x &=& r\gamma_2(1-z)y-\gamma_1 x\label{eq:x}\\
\dot y &=& \gamma_1 x-\gamma_2 y\label{eq:y}\\
\dot z &=& r\gamma_2(1-z)y\label{eq:z}
\end{eqnarray}

\begin{figure} 
\begin{center}
\includegraphics[width=15cm]{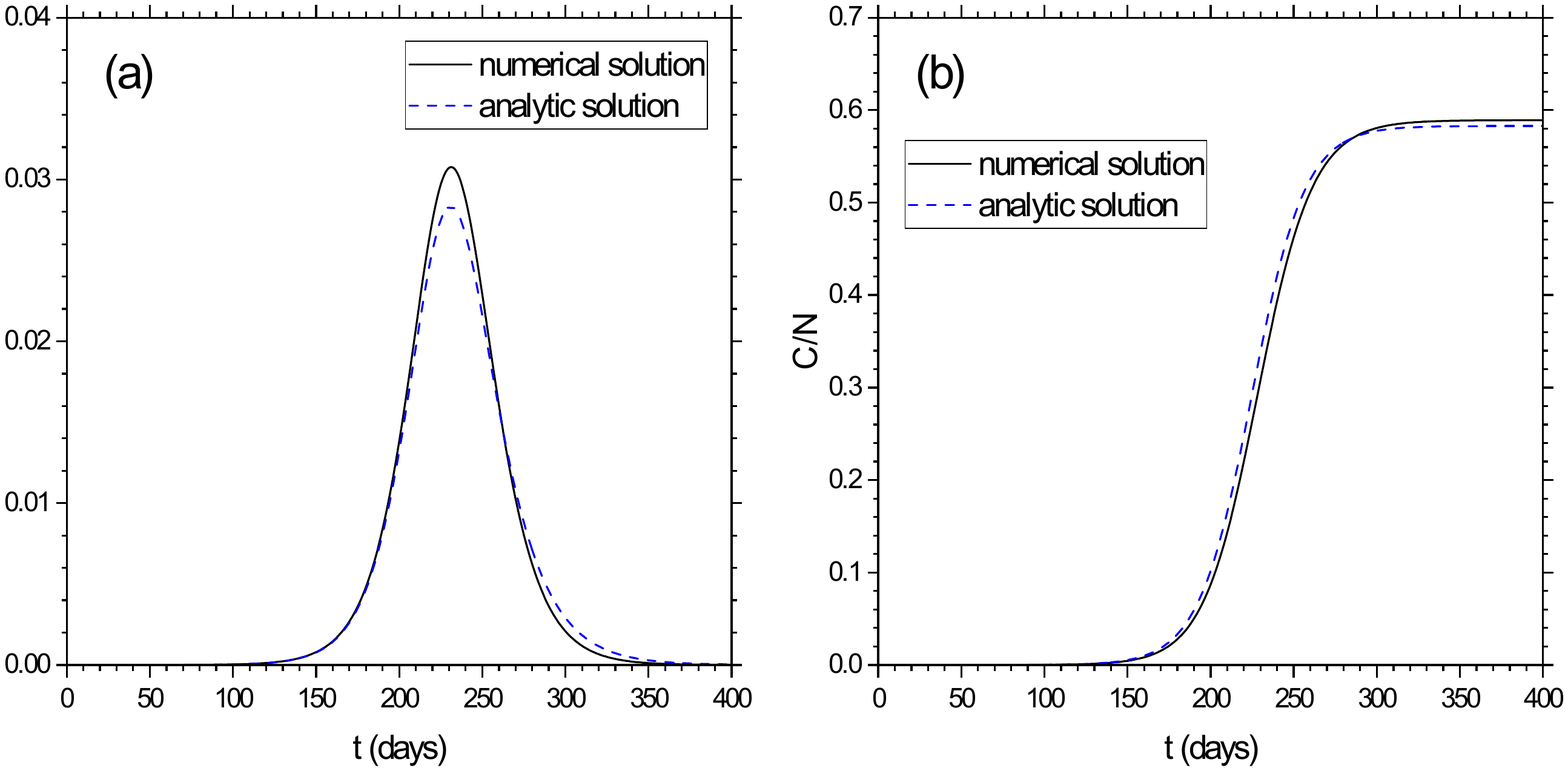}
\caption{Simulation with $r=1.5$ and $p=0.01$. Initial conditions: $E(0)=10$, $I(0)=0$, $N=6\cdot 10^7$. $I/N$ (a) and $C/N$ (b) vs. time from the numerical solution (solid black line) and from the analytic expressions, Eqs. (\ref{x:t}) and (\ref{z:t}) (dashed blue line). The time $t$ is in units of days and $\tau_E=3.69$ days, $\tau_I=3.48$ days.}
\label{Fig2}
\end{center}
\end{figure}

These equations have a single steady-state solution (i.e. $\dot x=\dot y=\dot z=0$) with $x=y=0$ (end of the epidemics) and $z=z_0$ with $0<z_0<1$. This solution is stable if $r_0=r(1-z_0)<1$. We see that the stability condition implies
\begin{equation}
z_0>1-\frac{1}{r}
\end{equation}
In Fig. \ref{Fig2} we plot $E/N$, $I/N$ and $C/N$ for $r=1.5$, $p=0.01$, $\tau_E=3.69$ days, $\tau_I=3.48$ days and initial conditions  $E(0)=10$, $I(0)=0$, $N(0)=6\cdot 10^7$. We observe that $C/N$ tends to a steady-state value of about $0.6$, whereas the peak of $I/N$ is about 0.03: it means that for these parameters the 60\% of the total population has contracted the infection and the peak the infected population is about 3\% of the total population. Note that these results are independent on $p$ and depend only on $\tau_E$, $\tau_I$ and $r$.

\subsection{Reduced model}

In this section we find an approximated analytic solution of Eqs.  (\ref{eq:x})-(\ref{eq:z}) in the free spread evolution with  $r>1$. The idea is to adiabatically eliminate the decaying mode with negative eigenvalue $\lambda_-$. To this aim, it is convenient to
write Eqs.(\ref{eq:x})-(\ref{eq:z}) in the basis of the eigenvalues $\lambda_\pm$. Writing again the linear equations (\ref{eq:E:lin}) and (\ref{eq:I:lin}) in the form
\begin{equation}
\frac{d}{dt}
\left(
\begin{array}{c}
  x \\
  y \\
\end{array}
\right)=
\left(%
\begin{array}{cc}
  -\gamma_1 & r\gamma_2 \\
   \gamma_1 & -\gamma_2 \\
\end{array}%
\right)\left(
\begin{array}{c}
  x \\
  y \\
\end{array}
\right)
\end{equation}
the normalized eigenvectors associated to the eigenvalues $\lambda_\pm$ of Eq.(\ref{eigen}) are
\begin{equation}
u_\pm=\frac{1}{D_{\pm}}\left(%
\begin{array}{c}
  \gamma_2+\lambda_\pm \\
  \gamma_1 \\
\end{array}%
\right)
\end{equation}
where
\begin{equation}
D_\pm=\sqrt{\gamma_1^2+(\gamma_2+\lambda_\pm)^2}
\end{equation}
Hence, in the new basis 
\begin{equation}
\left(
\begin{array}{c}
  x \\
  y \\
\end{array}
\right)=
\left(%
\begin{array}{cc}
  \frac{\gamma_2+\lambda_+}{D_+} & \frac{\gamma_2+\lambda_-}{D_-} \\
   \frac{\gamma_1}{D_+} & \frac{\gamma_1}{D_-} \\
\end{array}%
\right)\left(
\begin{array}{c}
  \bar x \\
  \bar y \\
\end{array}
\right)
\end{equation}
and the inverse is
\begin{equation}
\left(
\begin{array}{c}
  \bar x \\
  \bar y \\
\end{array}
\right)=
\left(%
\begin{array}{cc}
  \frac{\gamma_1}{D_-}  & -\frac{\gamma_2+\lambda_-}{D_-} \\
 -\frac{\gamma_1}{D_+}  & \frac{\gamma_2+\lambda_+}{D_+}  \\
\end{array}%
\right)\left(
\begin{array}{c}
  x \\
  y \\
\end{array}
\right)
\end{equation}
In the new basis Eqs.(\ref{eq:x})-(\ref{eq:z}) take the form:
\begin{eqnarray}
\dot{\bar x} &=& \lambda_+\bar x-\frac{r\gamma_1\gamma_2}{\sqrt{\Delta}}\left(\bar x+\frac{D_+}{D_-}\bar y\right)z\label{2:x}\\
\dot{\bar y} &=& \lambda_-\bar y+\frac{r\gamma_1\gamma_2}{\sqrt{\Delta}}\left(\frac{D_-}{D_+}\bar x+\bar y\right)z\label{2:y}\\
\dot z &=& r\gamma_1\gamma_2\left(\frac{\bar x}{D_+}+\frac{\bar y}{D_-}\right)(1-z)\label{2:z}
\end{eqnarray}
Notice that as expected in the linear regime the dynamics of $\bar x$ and $\bar y$ are uncoupled. Now we consider the free spread regime with $r>1$ such that $\lambda_+$ is positive and $\lambda_-$ is negative. If $r-1$ is small, then $|\lambda_-|\gg \lambda_+$ and we can adiabatically eliminate the 'slave' variable $\bar y$. Neglecting $\dot{\bar y}$ in (\ref{2:y}) we obtain
\begin{equation}
\frac{D_+}{D_-}\bar y\approx -\frac{r\gamma_1\gamma_2}{\sqrt{\Delta}}\frac{\bar x z}{\lambda_-+(r\gamma_1\gamma_2/\sqrt{\Delta})z}
\end{equation}
which when inserted in Eqs.(\ref{2:x}) and (\ref{2:z}) yields
\begin{eqnarray}
\dot{\bar x} &=& \frac{1}{\lambda_-+(r\gamma_1\gamma_2/\sqrt{\Delta})z}\left[\lambda_+\lambda_-+\frac{r\gamma_1\gamma_2}{\sqrt{\Delta}}(\lambda_+-\lambda_-)z\right]\bar x\label{3:x}\\
\dot z &=& \frac{r\gamma_1\gamma_2}{D_+}\left(\frac{\lambda_-}{\lambda_-+(r\gamma_1\gamma_2/\sqrt{\Delta})z}\right)\bar x(1-z)\label{3:z}
\end{eqnarray}
Since $\lambda_+-\lambda_-=\sqrt{\Delta}$ and $\gamma_1\gamma_2 r=\lambda_+\lambda_-[r/(1-r)]$,
\begin{eqnarray}
\dot{\bar x} &=& \frac{\lambda_+}{1-\beta z}\left[1-\frac{z}{k}\right]\bar x\label{4:x}\\
\dot z &=& -\frac{\lambda_+\lambda_-}{kD_+}\frac{(1-z)}{1-\beta z}\bar x\label{4:z}\\
\frac{D_+}{D_-}\bar y &=& \frac{\beta z}{1-\beta z}\bar x
\end{eqnarray}
where $k=(r-1)/r$ and $\beta=\lambda_+/k\sqrt{\Delta}$.
Finally, the original variables are
\begin{eqnarray}
y &=& \frac{\gamma_1}{D_+}\left[\frac{\bar x}{1-\beta z}\right]\label{y:def}\\
x &=& \frac{\gamma_2}{\gamma_1}\left[1+\frac{\lambda_+}{\gamma_1}\left(1-\frac{z}{k}\right)\right]y
\end{eqnarray}

\subsection{Analytical solution}

\begin{figure} 
\begin{center}
\includegraphics[width=10cm]{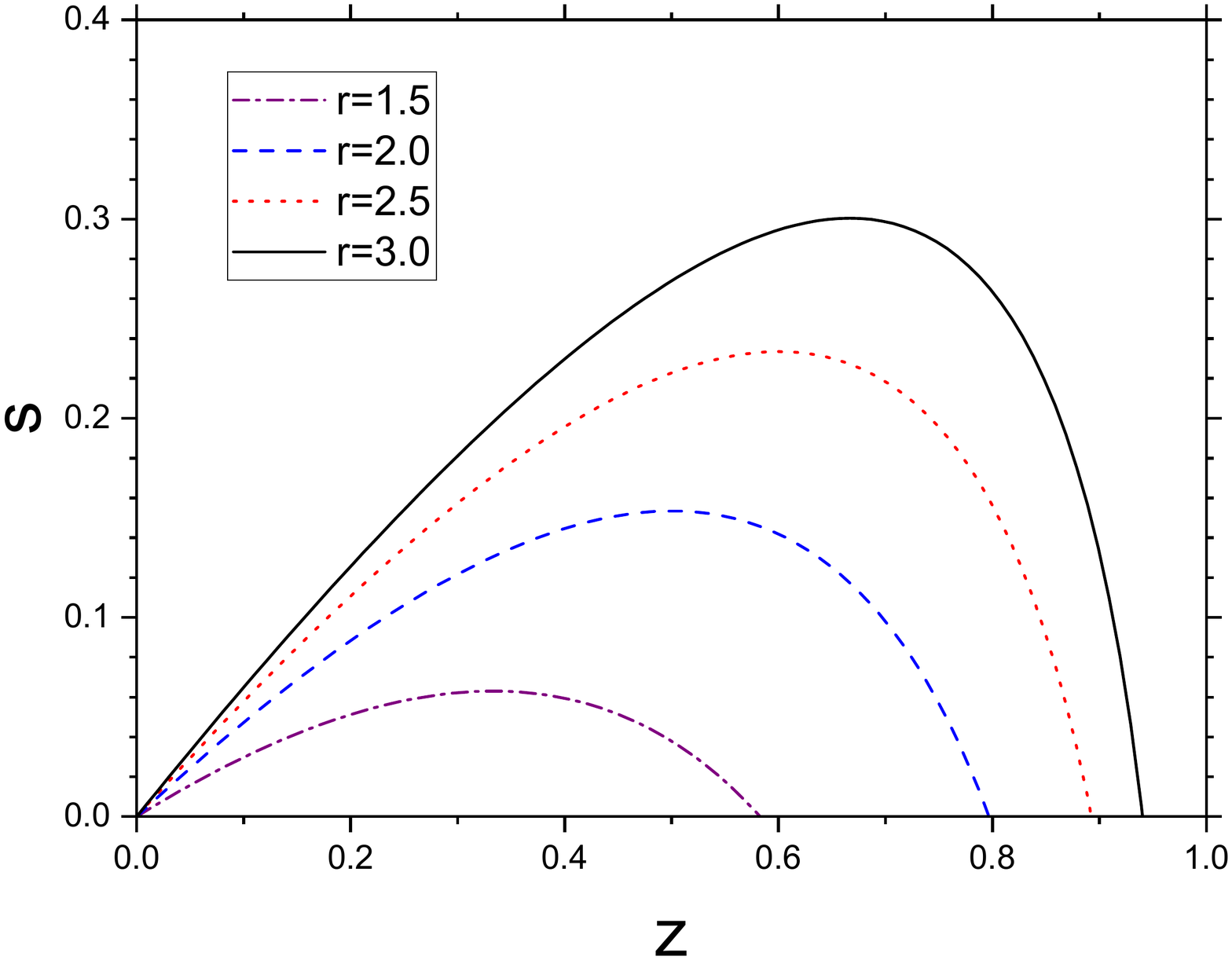}
\caption{Plot of $s$ vs. $z$ for $r=1.5,2,2.5,3.0$, from Eq.(\ref{s:z}).}
\label{s-vs-z}
\end{center}
\end{figure}

Eqs.(\ref{4:x}) and (\ref{4:z}) may provide some analytical result. Rescaling the time as
\begin{equation}
\tau=\frac{\lambda_+}{k}t
\end{equation}
and defining
\begin{eqnarray}
s &=& -\frac{\lambda_-}{D_+}\bar x\label{s:def}
\end{eqnarray}
Eqs.(\ref{4:x}) and (\ref{4:z}) take the form:
\begin{eqnarray}
\frac{ds}{d\tau}&=& \left(\frac{k-z}{1-\beta z}\right)s\label{s}\\
\frac{dz}{d\tau} &=&\left(\frac{1-z}{1-\beta z}\right)s\label{z}
\end{eqnarray}
In the limit $\beta\rightarrow 0$ they have the form of Lotka-Volterra equations \cite{brauer2012mathematical}. From them, dividing member by member, it results
\begin{eqnarray}
\frac{ds}{dz}&=& \frac{k-z}{1-z}
\end{eqnarray}
which when integrated yields
\begin{eqnarray}
s&=& z+\frac{1}{r}\ln|1-z|\label{s:z}
\end{eqnarray}
where we assumed $s\rightarrow 0$ when $z\rightarrow 0$. On the other hand, $s\rightarrow 0$ when $z\rightarrow z_\infty$ (see Fig.\ref{s-vs-z}), where $z_\infty$
is the solution of the transcendental equation
\begin{equation}
rz_\infty+\ln|1-z_\infty|=0\label{alpha}
\end{equation}
The same transcendental equation (\ref{alpha}) for $z_{\infty}$ has been obtained for the SIR compartmental model \cite{harko2014exact,miller2012note}. Here we have demonstrated its validity also for the SEIR model.

We see from Fig. \ref{s-vs-z} that $s=0$ for $z=0$ and $z=z_\infty$.
The maximum value of $s$ occurs when $z=k=1-1/r$ so that
\begin{equation}
s_{max}=1-\frac{1}{r}-\frac{1}{r}\ln r
\label{smax}
\end{equation}
These simple equations provide two analytic expressions for the asymptotic value of $C/N$ and for the peak of $I/N$. 

Let's now find an approximated solution of $z$ as a function of the scaled time $\tau$.
Using Eq. (\ref{s:z}) in Eq. (\ref{z}) we obtain a differential equation for $z$:
\begin{equation}
\frac{dz}{d\tau}=\frac{1-z}{1-\beta z}\left(z+\frac{1}{r}\ln|1-z|\right)
\end{equation}
From the numerical analysis and assuming $\beta z\ll 1$, we find that $z(\tau)$ is well approximated by the following function:
\begin{equation}
z(\tau)=\frac{z_\infty}{2}\left\{1+\tanh[k(\tau-\tau_d)/2]\right\}=\frac{z_\infty e^{k(\tau-\tau_d)}}{1+e^{k(\tau-\tau_d)}}\label{z:tau}
\end{equation}
where $\tau_d$ depends on the initial conditions. 
From (\ref{s}) it follows for $\beta z\ll 1$
\begin{eqnarray}
\frac{ds}{d\tau}&=& \left[k-z(\tau)\right]s
=\left\{k-\frac{z_\infty}{2}-\frac{z_\infty}{2}
\tanh[k(\tau-\tau_d)/2]\right\}s
\end{eqnarray}
This equation can be integrated to give
\begin{equation}
s(\tau)=s(0) \left\{
\frac{\cosh[k\tau_d/2]}{\cosh[k(\tau-\tau_d)/2]}\right\}^{z_\infty/k}e^{(k-z_\infty/2)\tau}\label{s:tau_1}
\end{equation}
Since $k\tau_d\gg 1$ and, from Eqs. (\ref{s:z}) and (\ref{z:tau}),  $s(0)\approx k z_\infty\exp(-k\tau_d)$, we can write Eq.(\ref{s:tau_1}) in the following form:
\begin{equation}
s(\tau)=kz_\infty\left[\frac{\mathrm{sech}[k(\tau-\tau_d)/2]}{2}\right]^{z_\infty/k}e^{(k-z_\infty/2)(\tau-\tau_d)}\label{s:tau:2}
\end{equation}
The time $\tau_{max}$ at which $s(\tau)$ is maximum can be evaluated from the condition $z(\tau_{max})=k$ which, using Eq. (\ref{z:tau}), yields
\begin{equation}
\tau_{max}=\tau_d+\frac{1}{k}\ln\left[\frac{k}{z_\infty-k}\right]
\end{equation}
where $\tau_d=(1/k)\ln[kz_\infty/s(0)]$.
For instance, for $r=1.5$ and $s(0)=10^{-5}$, we obtain $z_\infty=0.5828$, $s_{max}=0.063$ and $\tau_{max}=31.25$.

\section{Results and Conclusions}

We have obtained analytical expressions for the asymptotic value of the cumulative population fraction $C/N$ and the peak of the infectious population fraction $I/N$ in the case of free spread evolution of COVID-19. Furthermore, we have obtained approximated expressions of these quantities as a function of time and the times at which the peak and the end of the epidemics is expected. We summarize here below these results:
\begin{figure} 
\begin{center}
\includegraphics[width=15cm]{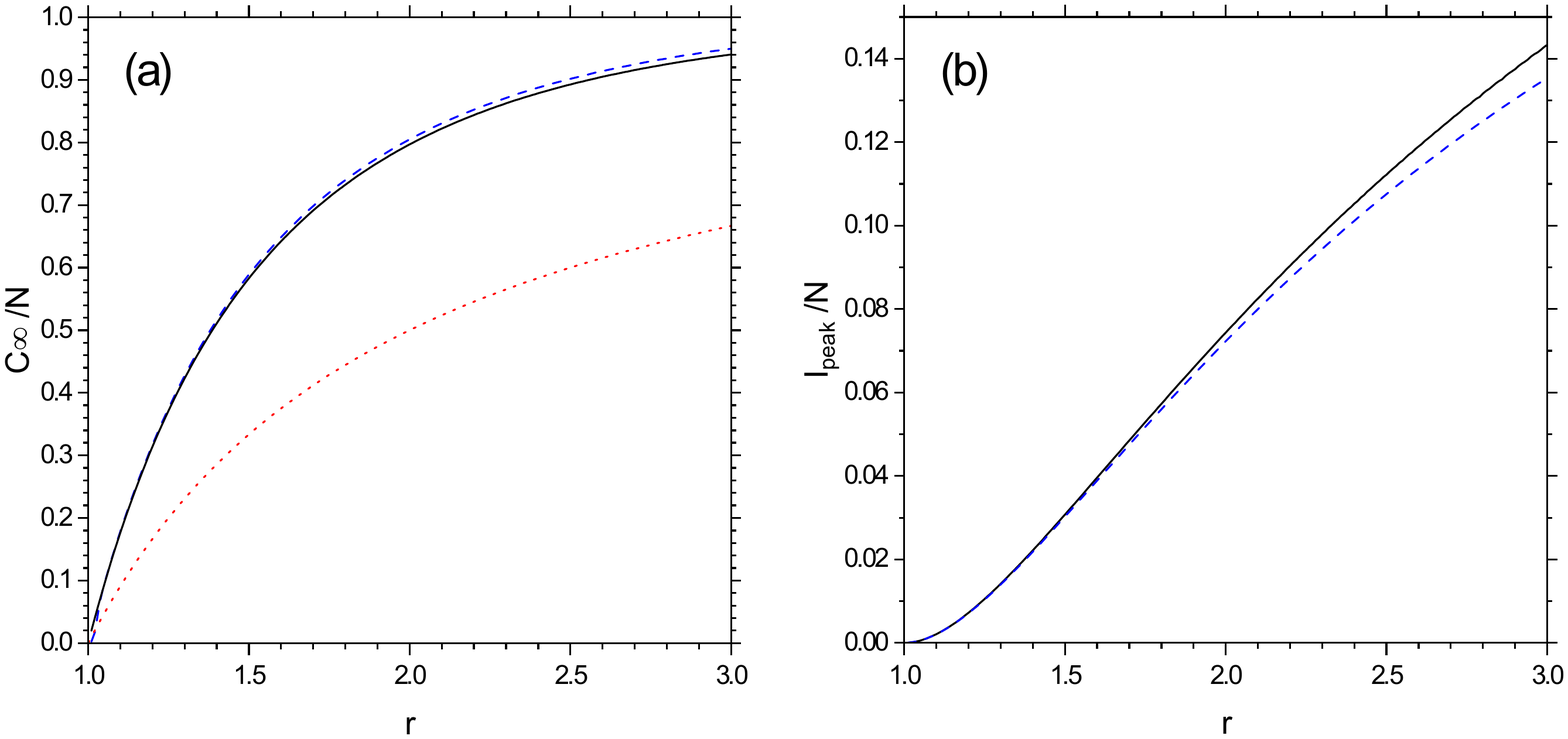}
\caption{(a): Plot of $C_\infty/N$ vs. $r$, from the numerical solution of Eqs.(\ref{eq:S})-(\ref{eq:D}) (dashed line) and from the analytical result of Eq.(\ref{alpha}) (continuous line). The dotted red line is the threshold value $k=1-1/r$. (b) Peak value of $I/N$ vs. $r$ from the numerical solution of Eqs.(\ref{eq:S})-(\ref{eq:D}) (dashed line) and from  Eq.(\ref{Ipeak}) (continuous line)}\label{Fig4}
\end{center}
\end{figure}
\begin{description}
\item[(a)] The asymptotic value of the cumulative population fraction is $C_{\infty}/N=z_\infty$, where $z_\infty$ is the solution of the transcendental equation (\ref{alpha}). A comparison between the exact solution obtained by integrating Eqs.(\ref{eq:S})-(\ref{eq:D}) and the solution of Eq.(\ref{alpha}) is shown in Fig.\ref{Fig4}(a). Notice that this value depends only on the reproduction number $r$.
\item[(b)] The peak value of the infectious population fraction is, from Eqs.(\ref{y:def}),(\ref{s:def}) and (\ref{smax}),
\begin{equation}
\frac{I_{\mathrm{peak}}}{N}=\frac{4\gamma_1\sqrt{\Delta}}{(\gamma_1+\gamma_2+\sqrt{\Delta})^2}
\left[1-\frac{1}{r}-\frac{1}{r}\ln r\right]\label{Ipeak}
\end{equation}
The agreement of this expression with the exact result shown in Fig.\ref{Fig4}(b) is better for  values of $r$ closer to the threshold $r=1$.
\item[(c)] We have obtained an approximated temporal profile of $C(t)/N$,
\begin{equation}
\frac{C(t)}{N}=z(t)=\frac{z_\infty}{2}\left\{1+\tanh[\lambda_+(t-t_d)/2]\right\}\label{z:t}
\end{equation}
where $t_d=(1/\lambda_+)\ln[kz_\infty/s_0]$ and $s_0=(-\lambda_-/D_+D_-)[\gamma_1 x_0-(\gamma_2+\lambda_-)y_0]$, where $x_0$ and $y_0$ are the initial values of $x$ and $y$. From this expression we have obtained the expression of $I(t)/N$ as a function of time:
\begin{equation}
\frac{I(t)}{N}=\left(-\frac{\gamma_1}{\lambda_-}\right)\frac{s(t)}{1-\beta z(t)}\label{x:t}
\end{equation}
where $\beta=\lambda_+/k\sqrt{\Delta}$ and
\begin{equation}
s(t)=kz_\infty\left[\frac{\mathrm{sech}[\lambda_+(t-t_d)/2]}{2}\right]^{z_\infty/k}e^{\lambda_+(1-z_\infty/2k)(t-t_d)}\label{s:t}
\end{equation}
The good agreement of Eqs.(\ref{z:t}) and (\ref{x:t}) with the exact numerical solution of Eqs.(\ref{eq:S})-(\ref{eq:D}) is shown in Fig. \ref{Fig2}.
\item[(d)] The time at which the peak of $I/N$ is reached is
\begin{equation}
t_{\mathrm{peak}}=\frac{1}{\lambda_+}\ln\left[\frac{k^2z_\infty}{s_0(z_\infty-k)}\right]
\end{equation}
Fig. \ref{Fig5} shows $t_{\mathrm{peak}}$ (in units of days) as a function of $r$ for an initial value of $E(0)=10$, $I(0)=0$ and $N(0)=6\cdot 10^7$.
\end{description}
\begin{figure} 
\begin{center}
\includegraphics[width=12cm]{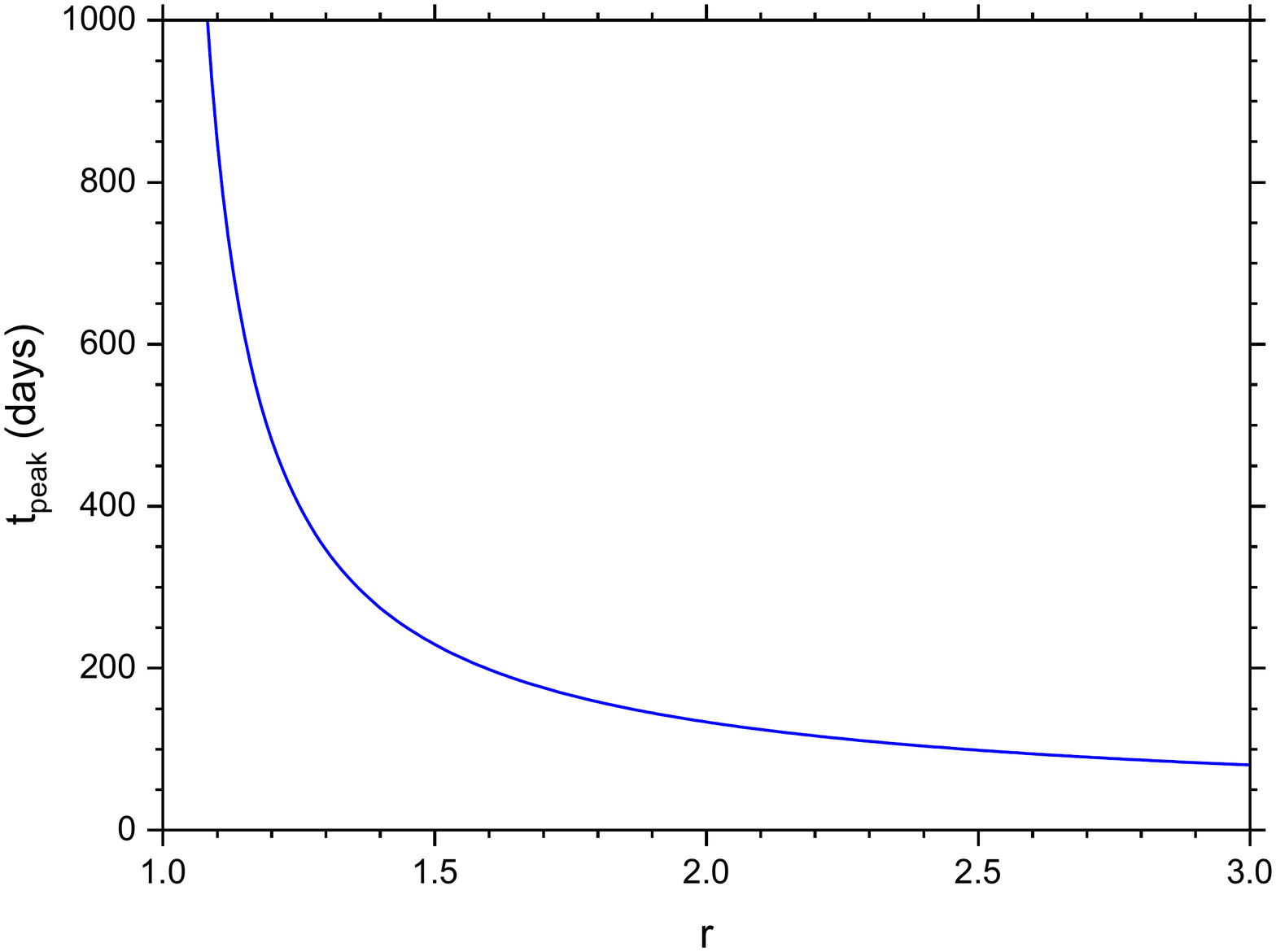}
\caption{Plot of peak time $t_{\mathrm{peak}}$ (in units of days) vs. $r$ for initial values of $E(0)=10$ and $I(0)=0$, $N(0)=6\cdot 10^7$ and $\tau_E=3.69$ days, $\tau_I=3.48$ days.}\label{Fig5}
\end{center}
\end{figure}
These analytic expressions can be useful for deriving the uncertainty in the estimates of COVID-19 caused by the fluctuations of the values of the control parameters, as for instance the reproduction number $r$. In fact, the results of ref.\cite{faranda2020modelling} suggest that uncertainties in both parameters and initial conditions rapidly propagate in the model and can result in different outcomes of the epidemics. For instance, Fig.\ref{Fig4}a and \ref{Fig4}b show the dependence of the fraction of the final cumulative fraction, $C_{\infty}/N$, and the daily infections peak, $I_{\mathrm{peak}}/N$, as a function of $r$. We observe that the sensitivity of $C_{\infty}/N$  on $r$ variations is larger when $r$ is close to unity (with approximately $C_{\infty}/N\approx 2(r-1)$) whereas it decreases for increasing values of $r$. On the other hand, $I_{\mathrm{peak}}/N$ grows almost linearly with $r$ (approximately as $I_{\mathrm{peak}}/N\approx 0.07(r-1)$), so that its sensitivity to $r$ variations is almost constant. Finally, the uncertainty of the peak time $t_{\mathrm{peak}}$ (see fig.5) on $r$ variations is very large for $r$ close to unity and it reduces strongly at larger $r$.

In conclusions, we have obtained analytical expressions for the peak and asymptotic values of  COVID-19 pandemic curves in the free spread as a function of the reproduction number and the two average times in the exposed and infected states. The results have been obtained by reducing the exact nonlinear model by adiabatically eliminating the decaying mode of the linear regime. This allows to reduce the SEIR model of a set of two equations similar to the Lotka-Volterra equations, from which exact and approximated solutions can be obtained. The analytical results have been compared with the exact numerical solution, showing good agreement. Particular interesting is the asymptotic fraction of the removed (recovered+deaths) population fraction, which depends only on the reproduction number $r$. Finally, the infected population curve is an almost symmetric function described by an hyperbolic secant function.

\bibliographystyle{elsarticle-num} 
\bibliography{references_Covid}

\end{document}